\documentclass[11pt]{article}
\usepackage{mydef2col}
\usepackage{kantlipsum,widetext}
\usepackage[T1]{fontenc}
\usepackage[autostyle=true]{csquotes}

\allowdisplaybreaks[3]
\usepackage{tikz}
\usetikzlibrary{calc,decorations.markings}
\usetikzlibrary{arrows,patterns,positioning}

\usepackage[round]{natbib}

\def\baru{{\bar{u}}}

\newcommand{\iu}{\mathrm{i}\mkern1mu}

\DeclareMathOperator{\erf}{erf}

\title{Semi-closed form solutions for barrier and American options  written on a time-dependent Ornstein Uhlenbeck process }
\def\thetitle1{Semi-closed form solutions for barrier and American options ...}
\author{
\authorstyle{Peter Carr
and Andrey Itkin{}
}
\newline\newline
\institution{Tandon School of Engineering, New York University, 1 Metro Tech Center, 10th floor, Brooklyn NY 11201, USA}}

\date{\today}
\begin{document}

\maketitle

\lettrineabstract{In this paper we develop a semi-closed form solutions for the barrier (perhaps, time-dependent) and American options written on the  underlying stock which follows a time-dependent OU process with a lognormal drift. This model is equivalent to the familiar Hull-White model in FI, or a time dependent OU model in FX. Semi-closed form means that given the time-dependent interest rate, continuous dividend and volatility functions, one need to solve numerically a linear (for the barrier option) or nonlinear (for the American option) Fredholm equation of the first kind. After that the option prices in all cases are presented as one-dimensional integrals of combination of the above solutions and Jacobi theta functions. We also demonstrate that computationally our method is more efficient than the backward finite difference method used for solving these problems, and can also be as efficient as the forward finite difference solver while providing better accuracy and stability.}

\vspace{0.5in}

\section*{Introduction}

The Orstein-Uhlenbeck (OU) process with time-dependent coefficients is very popular among practitioners for modeling interest rates and credit. That is because it is relatively simple, allows negative interest rates (which recently has become a hot feature) , and could be calibrated to the given term-structure of interest rates and to the prices or implied volatilities of caps, floors or European swaptions since the mean-reversion level and volatility are functions of time. The most known among this class are Hull-White and Vasicek models, see \citep{BM2006} and references therein.

The Hull-White model is a one-factor model for the stochastic short interest rate $r_t$ of the form
\begin{equation} \label{OU}
d r_t = k[\theta(t) - r_t] dt + \sigma(t)dW_t,
\end{equation}
\noindent where $t$ is the time, $k > 0$ is the constant speed of mean-reversion, $\theta(t)$ is the mean-reversion level, $\sigma(t)$ is the volatility of the process, $W_t$ is  the standard Brownian motion under the risk-neutral measure. This model can also be used for pricing Equity or FX derivatives if one assumes that the mean-reversion level vanishes, while the mean-reversion rate is replaced either by $q(t) - r(t)$ for Equities, or by $r_f(t) - r_d(t)$ for FX, where $r(t), q(t)$ are the deterministic interest rate and continuous dividends, and $r_d(t), r_f(t)$ are the deterministic domestic and foreign interest rates.

Without loss of generality in this paper we are concentrated on the Equity world. Since the process in \eqref{OU} is Gaussian the model is tractable for pricing European plain vanilla options. However, for exotic options, e.g., highly liquid barrier options, or for American options these prices are not known yet in closed form. Therefore, various numerical methods are used to obtain them, that sometimes could be computationally expensive. In this paper we attack this problem by constructing a semi-closed form solutions for the prices of barrier and American options written on the process \eqref{OU}. The results obtained in the paper are new. Our approach to a certain degree is similar to that in \citep{Mijatovic2010}, who, however, used a different underlying process (the lognormal model with local spot-dependent volatility,
and constant interest rates and dividends, but time-dependent barriers). Therefore, our model is more general in a sense that all parameters of the model are time-dependent, while adding time-dependent barriers can be naturally done within our approach. Also, as compared with \citep{Mijatovic2010}, we don't use a probabilistic argument, rather a theory of partial differential equations (PDE). At the end we demonstrate that computationally our method is more efficient than the backward finite difference method used to solve these problems, and can also be as efficient as the forward finite difference solver while providing better accuracy and stability.

\section{Problem for pricing barrier options} \label{SecBO}

We start by specifying the dynamics of the underlying stock price $S_t$ to be
\begin{equation} \label{OU1}
d S_t = [r(t) - q(t)] S_t dt + \sigma(t)dW_t, \qquad S_{t=0} = S_0,
\end{equation}
\noindent where now $r(t)$ is the deterministic short interest rate, and $S_t$ is the stock price\footnote{It is easy to show that this model is equivalent to the Hull-White model.}. Here we don't specify the explicit form of $r(t, q(t), \sigma(t)$ but assume that they are known either as a continuous functions of time $t \in [0,\infty)$, or as a discrete set of $N$ values for some moments $t_i, \ i=1,\ldots,N$.

Further in this section we consider a contingent claim written on the underlying process $S_t$ in \eqref{OU1} which is the Up-and-Out barrier Call option. It is known that by the Feynman-Kac formula, \citep{klebaner2005} one can obtain a parabolic ({\it linear}) PDE which solution gives the Up-and-Out barrier Call option price $C(S,t)$ conditional on $S_{t=0} = S$, which reads
\begin{equation} \label{PDE}
\fp{C}{t} + \dfrac{1}{2}\sigma^2(t) \sop{C}{S} +  [r(t) - q(t)] S \fp{C}{S} = r(t) C.
\end{equation}
This equation should be solved subject to the terminal condition at the option maturity $t=T$
\begin{equation} \label{tc0}
C(S,T) = (S-K)^+,
\end{equation}
\noindent and the boundary conditions
\begin{equation} \label{bc0}
C(0,t) = 0, \qquad C(H,t) = 0,
\end{equation}
\noindent where $H$ is the upper barrier.

Our goal now is to build a series of transformations to transform \eqref{PDE} to the heat equation.

\subsection{Transformation to the heat equation} \label{trHeat}

To transform the PDE \eqref{PDE} to the heat equation we first make a change of the dependent and independent variables as follows:
\begin{equation} \label{tr1}
S \to x/g(t), \qquad C(S,t) \to e^{f(x,t)}u(x,t),
\end{equation}
\noindent where new functions $f(x,t), g(t)$ has to be determined in such a way, that the equation for $u$ is the heat equation. This can be done by substituting \eqref{tr1} into \eqref{PDE} and providing some tedious algebra. The result reads
\begin{align} \label{sub1}
f(x,t) & =k(t)  - \frac{g'(t)+g(t) (r(t)-q(t))}{2 g(t)^3 \sigma (t)^2} x^2 , \\
k(t) &= \frac{1}{2} \log \left(\frac{g(t)}{g(0)}\right) + \frac{1}{2} \int_0^t [3 r(s)-q(s)]  ds. \nonumber
\end{align}
The function $g(t)$ solves the following ordinary differential equation (ODE)
\begin{align} \label{gODE}
0 &= b(t) g(t)-g''(t) + 2 g'(t) \frac{\sigma '(t)}{\sigma (t)}+ 2 \frac{g'(t)^2}{g(t)}, \\
b(t) &= 2 (r(t) - q(t)) \frac{\sigma'(t)}{\sigma(t)} - \left[ (r(t)-q(t))^2 + r'(t) - q'(t) \right]. \nonumber
\end{align}
The \eqref{gODE} by substitution
\begin{equation} \label{gDef}
g(t) \to e^{\int_0^t w(s) ds}
\end{equation}
\noindent  can be further transformed to the Riccati equation
\begin{equation} \label{Ric}
w'(t) = b(t) + w(t)^2  + 2 w(t) \frac{\sigma '(t)}{\sigma (t)}.
\end{equation}
This equation cannot be solved analytically for arbitrary functions $r(t), q(t), \sigma(t)$, but can be efficiently solved numerically. Also, in some cases it can be solved in closed form. For instance, if $|r(t) - q(t)| t  = \varepsilon \ll 1$  (which at the current market is a typical case), then $b(t)$ can be reduced to $b(t) = 2 (r(t) - q(t)) \sigma'(t)/\sigma(t)$. Then assuming in the first approximation on $\varepsilon$
\begin{equation} \label{bsmall}
|w(t)| \gg |r(t) - q(t)|,
\end{equation}
\noindent we obtain the solution
\begin{equation} \label{w}
w(t) = \frac{\sigma^2(t)}{D - \int_0^t \sigma^2(s) ds},
\end{equation}
\noindent where $D$ is an integration constant. Thus, \eqref{bsmall} can be re-written as
\begin{equation*} \label{w1}
\Var(t) \left[1 +  \frac{V}{\bar{V}} \varepsilon\right] \gg D (r(t)-q(t)),
\end{equation*}
\noindent where $\Var(t) = \sigma^2(t)$ is the normal variance, and $\bar{V}(t) = \frac{1}{t}\int_0^t \sigma^2(s) ds$ is the average normal variance. Thus, our solution in \eqref{w} is correct if $\varepsilon \ll 1$ and also $V \varepsilon/\bar{V} \ll 1$, because then $D$ can always be chosen to obey the inequality $\Var(t) \gg D (r(t)-q(t)), \ \forall t \in [0,T]$.

With these explicit definitions \eqref{PDE} transforms to the form
\begin{equation} \label{PDE1}
\frac{1}{2} \sigma (t)^2 e^{2 \int_0^t w(s) \, ds} \sop{u}{x} + \fp{u}{t} = 0.
\end{equation}
The next step is to make a change of the time variable
\begin{equation} \label{tTr2}
\tau(t) \to  \frac{1}{2} \int_t^T \sigma^2(s) e^{2 \int_0^s w(m) \, dm} \, ds,
\end{equation}
\noindent so \eqref{PDE1} finally takes the form of the heat equation
\begin{equation} \label{Heat}
\fp{u}{\tau} = \sop{u}{x}.
\end{equation}
The \eqref{Heat} should be solved subject to the terminal condition
\begin{equation} \label{tc}
u(x,0) = (x e^{-\int_0^T w(s) ds} - K)^+ e^{-f(x,T)},
\end{equation}
\noindent and the boundary conditions
\begin{align} \label{bc}
u(0,\tau) &= 0, \qquad u(y(\tau),\tau) = 0, \qquad y(\tau) = H g(t(\tau)).
\end{align}
These conditions directly follow from \eqref{tc0}, \eqref{bc0}, while $y(\tau)$ is now a time-dependent upper barrier
\footnote{Therefore, we can also naturally solve the same problem with the time-dependent upper barrier $H=H(t)$ as this just changes the definition of $y(\tau)$.}. The function $t(\tau)$ is the inverse map of \eqref{tTr2}. It can be computed for any $t \in [0,T]$ by substituting it into \eqref{tTr2}, then finding the corresponding value of $\tau(t)$, and finally inverting.

\subsection{Solution of the barrier pricing problem} \label{GITmethod}

The PDE in \eqref{Heat}, \eqref{tc}, \eqref{bc} is a parabolic equation which solution should be found at the domain with moving boundaries. These kind of problems are known in physics for a long time. Similar  problems arise in the field of nuclear power engineering and safety of nuclear reactors; in studying combustion in solid-propellant rocket engines; in  laser  action  on  solids; in the theory of phase transitions (the  Stefan  problem  and  the  Verigin  problem  (in  hydromechanics)); in the processes of sublimation in freezing and melting; in  the  kinetic  theory  of  crystal  growth; etc., see \citep{kartashov1999} and references therein. Analytical solutions of these problems require non-traditional, and sometimes  sophisticated methods. Those methods were actively elaborated on by the Russian mathematical school in the 20th century starting from A.V.~Luikov, and then by B.Ya.~Lyubov, E.M.~Kartashov, and many others.

As applied to mathematical finance, one of these methods - the method of heat potentials - was actively utilized by A.~Lipton and his co-authors who solved various problems of mathematical finance by using this approach, see \citep{Lipton2002,LiptonPrado2020} and references therein. Another method that we use in this paper is the method of a generalized integral transform. Below we closely follow \citep{kartashov2001} when give an exposition of the method.

 We start by introducing an integral transform of the form
 \begin{equation} \label{intTr}
 \bar{u}(p,\tau) = \int_0^{y(\tau)} u(x,\tau) \sinh(x \sqrt{p}) dx,
 \end{equation}
 \noindent where $ p = a + \iu \omega$ is a complex number with $\mathds{R}(p) \ge \beta > 0$, and $- \frac{\pi}{4} < \arg\left(\sqrt{p}\right) < \frac{\pi}{4}$. Let us multiply both parts of \eqref{Heat} by $\sinh(x \sqrt{p})$ and then integrate on $x$ from zero to $y(\tau)$:
 \begin{equation*} \label{H2}
\int_0^{y(\tau)} \fp{u}{\tau}\sinh(x \sqrt{p}) dx = \int_0^{y(\tau)} \sop{u}{x} \sinh(x \sqrt{p}) dx.
\end{equation*}
Integrating by parts, we obtain
 \begin{align*} 
\fp{}{\tau} & \int_0^{y(\tau)}u(x,\tau)\sinh(x \sqrt{p}) dx - u(y(\tau),\tau) \sinh(y(\tau) \sqrt{p}) y'(\tau) \\
&= \fp{u(x,\tau)}{x} \sinh(x \sqrt{p})\Bigg|_{0}^{y(\tau)} + \sqrt{p} u(x,\tau) \cosh(x \sqrt{p})\Bigg|_{0}^{y(\tau)} +
p \int_0^{y(\tau)}u(x,\tau)\sinh(x \sqrt{p}) dx. \nonumber
\end{align*}
With allowance for the boundary conditions in \eqref{bc} and the definition in \eqref{intTr} we obtain the following Cauchy problem
\begin{align} \label{ode2}
\fp{\baru}{\tau} &- p \baru = \Psi(\tau) \sinh(y(\tau) \sqrt{p}),  \\
\baru(p,0) &=  \int_0^{y(0)} u(x,0) \sinh(x \sqrt{p}) dx, \qquad \Psi(\tau) = \fp{u(x,\tau)}{x}\Big|_{x = y(\tau)}.
\nonumber
\end{align}
 The \eqref{ode2} can be solved explicitly, assuming that $\Psi(\tau)$ is known. The solution reads
\begin{equation} \label{sol1}
\baru e^{- p \tau}  = \int_0^\tau \Psi(k) e^{- p k} \sinh(y(k) \sqrt{p}) d k
 +  \int_0^{y(0)} u(x,0) \sinh(x \sqrt{p}) dx.
\end{equation}
 As $\mathds{R}(p) \ge \beta > 0$, and $\baru(x,\tau) < \infty$, the function $\baru e^{- p \tau} \to 0$ at $\tau \to \infty$. Therefore, letting $\tau$ tend to $\infty$, we obtain an equation which makes connection between the moving boundary $y(\tau)$ and $\Psi(\tau)$:
 \begin{equation} \label{sol2}
\int_0^\infty \Psi(\tau) e^{- p \tau} \sinh(y(\tau) \sqrt{p}) d \tau = -\int_0^{y(0)} u(x,0) \sinh(x \sqrt{p}) dx.
\end{equation}
Using the definitions in \eqref{tc} and \eqref{sub1}, the integral in the RHS of \eqref{sol2} can be represented as
\begin{align} \label{Fp}
F(p) &\equiv - e^{-\int_0^T w(s) ds}  \int_{K_1}^{y(0)} (x - K_1) \sinh(x \sqrt{p})  e^{k (T) - a(T) x^2} dx \\
&= e^{-\int_0^T w(s) ds}  \frac{e^{k(T) -x \left[a(T) x+\sqrt{p}\right]}}{8 a^{3/2}(T)}
\Bigg\{2 \sqrt{a(T)} \left(e^{2 \sqrt{p} x} - 1 \right) -  \sqrt{\pi } e^{\frac{\left[2 a(T) x+\sqrt{p}\right]^2}{4 a(T)}} \nonumber \\
&\cdot \left[ \left[\sqrt{p}-2 a(T) K_1\right] \erf \left(\frac{2 a(T) x-\sqrt{p}}{2 \sqrt{a(T)}} \right)
+ \left[\sqrt{p} + 2 a(T) K_1 \right] \erf \left(\frac{2 a(T) x+\sqrt{p}}{2 \sqrt{a(T)}}\right) \right]
\Bigg\}\Bigg|_{K_1}^{y(0)}, \nonumber \\
K_1 &= K e^{\int_0^T w(s) ds}, \quad a(t) = \frac{g'(t)+g(t) (r(t)-q(t))}{2 g(t)^3 \sigma (t)^2}. \nonumber
\end{align}
 Thus, \eqref{sol2} takes the form
 \begin{equation} \label{fred}
\int_0^\infty \Psi(\tau) e^{- p \tau} \sinh(y(\tau) \sqrt{p}) d \tau = F(p),
\end{equation}
\noindent where F(p) is known from \eqref{Fp}.

The \eqref{fred} is a linear Fredholm integral equation of the first kind, \citep{polyanin2008handbook}. The solution $\Psi(\tau)$ can be found numerically on a grid by solving a system of linear equations. In other words, given functions $r(t), q(t), \sigma(t)$ we can compute first $w(t)$, then $g(t)$, and finally $\tau(t)$ (or $t(\tau))$, thus determining the moving boundary $y(\tau)$. Next we can solve \eqref{fred} for $\Psi(\tau)$ and substitute it into \eqref{sol1} to obtain the generalized transform of $u(x,\tau)$ in the explicit form. Therefore, if this transform can be inverted back, we solved the problem of pricing Up-and-Out barrier Call options.

 \subsection{The inverse transform}

In this Section the description of inversion is borrowed from \citep{kartashov2001}. Since that book has never been translated into English, we provide a wider exposition of the method. Also, the book contains various typos that are fixed here.

 As known from a general theory of the heat equation, the solution of the heat equation ${\cal L}u(x,\tau ) = 0, \ {\cal L} = \partial/\partial \tau - \nu\partial^2/\partial x^2$ at the space domain $0 < x < l = const$ can be expressed via Fourier series of the form, \citep{Polyanin2002}
\begin{equation*}\label{Four}
u(x,\tau ) = \sum_{n=1}^{\infty} \alpha_n e^{-\nu \gamma_n^2 t} \sin\left(\frac{n\pi x}{l}\right),
\end{equation*}
\noindent where $\psi (x) = \sin(n\pi x/l)$ are the eigenfunctions of the heat operator ${\cal L}$, and $\gamma _n = n\pi/l$ are its eigenvalues.

Therefore, by analogy we look for the inverse transform of $\baru$, or for the solution of \eqref{sol1} in terms of $u(x, \tau)$, to be a generalized Fourier series of the form, \citep{kartashov2001}
\begin{equation} \label{WFour}
u(x,\tau) = \sum_{n=1}^{\infty} \alpha_n(\tau ) e^{ - \left( \frac{n \pi}{y(\tau)}\right)^2 \tau } \sin\left(\dfrac{n\pi x}{y(\tau)} \right).
\end{equation}
\noindent where $\alpha(\tau)$ are some functions to be determined.  Note, that this definition automatically respects the vanishing boundary conditions for $u(x,\tau )$. We assume that this series converges absolutely and uniformly $\forall x \in [0,y(\tau)]$ for any $\tau > 0$.

\begin{figure}[!ht]
\hspace{-0.5in}
\begin{minipage}{0.6\textwidth}	
\captionsetup{width=0.8\linewidth}	
\begin{center}
\fbox{
\begin{tikzpicture}[thick, scale=0.5]
\def\axisXlength{8}
\def\zero{1}
\def\gammap{3}
\def\step{0.35}
\def\alpha{0.75}
\def\shif{-0.45}
\draw (-\axisXlength, 0) -- (0.5*\axisXlength,0)
      (\zero, -\axisXlength) -- (\zero, \axisXlength);

\foreach \i in {1,...,4} {
\draw[red, ultra thick, decoration={ markings,
      mark=at position 0.4 with {\arrow{latex}},
      mark=at position 0.52 with {\arrow{latex}},
      mark=at position 0.8 with {\arrow{latex}},
      mark=at position 0.97 with {\arrow{latex}},
	},
      postaction={decorate}]
	let
		\n1 = {\i*\i*\step + \i*\step + 0.5*\step}
	in
		(\gammap,-\n1) -- (\gammap,\n1) -- (\zero, \n1) arc (90:270:\n1)   --  (\zero, -\n1)  -- cycle;

    \node at (\zero - \i*\i*\step,0) {$\bullet$};
    \node at (\zero - \i*\i*\step,\shif) {$p_\i$};
    \node at (\zero - \i*\i*\step - \i*\step - 0.5*\step - 0.2, \i*\i*\step*\alpha + \i*\step*\alpha + 0.5*\step*\alpha - 0.07*\i*\i) {$\gamma_\i$};
}
\node at (0.6*\axisXlength,\shif){$\operatorname{Re} p$};
\node at (\zero,\axisXlength-\shif) {$\operatorname{Im} p$};
\node at (1.1*\gammap,\shif){$\gamma$};
\node at (1.4*\zero,\shif){$0$};
\node at (\zero - 25*\step,\shif) {$\ldots$};
\end{tikzpicture}
}
\end{center}
\caption{Contours of integration in a complex plane: $p_n, \ n=1,2...$ are simple poles of the LHS of the \eqref{WInver2}, $\gamma _n$ - the integration contours.}
\label{contours2}
\end{minipage}
\hspace{-0.2in}
\begin{minipage}{0.4\textwidth}
\centering
\includegraphics[totalheight=2.7in]{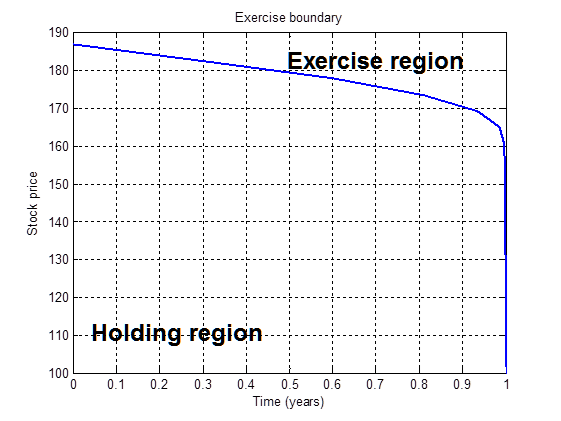}
\captionsetup{width=0.5\textwidth, margin={0.5in,0in}}	
\captionof{figure}{\hfill Typical exercise boundary for the American Call option.}
\label{EB}
\end{minipage}
\end{figure}

Applying this generalized integral transform to both parts of \eqref{intTr} and integrating, we obtain
\begin{equation} \label{WInver2}
 \sum_{n=1}^{\infty} \dfrac{(-1)^{n+1}n \alpha_n(\tau ) e^{ - \left( n \pi/y(\tau) \right)^2 \tau }}
{p + (n \pi/y(\tau))^2} = \dfrac{y(\tau)}{\pi \sinh\left(\sqrt{p} y(\tau)\right)} \baru(p,\tau ).
\end{equation}
The LHS of this equation is regular everywhere except simple poles on the negative semi-axis, see Fig,~\ref{contours2}
\begin{equation*}\label{poles1}
 p_n = - \left( \dfrac{n \pi}{y(\tau)} \right)^2, \quad n=1,2,...
\end{equation*}
Let us  sequentially integrate both sides of \eqref{WInver2} on $p$ along contours $\gamma_1, \gamma_2,\ldots$. The contour $\gamma _n$ consists of the vertical line $\gamma > 0$, the half-round of radius $R_n = [\pi^2/(2 y(\tau)](2n^2+2n+1)$ (the contour $\gamma_n$ crosses the $Re(p)$ axis in the middle point between $p_n$ and $p_{n+1}$ with the center in the origin), and two horizontal lines $Y=\pm [\pi^2/(2 y(\tau))](2n^2+2n+1)$. It means that the circle $R_n$ doesn't hit any pole of the LHS of \eqref{WInver2}. Then by the Cauchy's residual theorem, \citep{cauchy1984} the integral taken along the contour $\gamma _n$ is equal to $2\pi \iu$ times the sum of residuals of the LHS of \eqref{WInver2} that lie inside $\gamma _n$

As poles are simple, and the function under the integral in the LHS of \eqref{WInver2} has the form $F_1(p)/F_2(p)$, the residual of such a function is, \citep{cauchy1984}
\begin{equation*}\label{rs_form}
  \mathrm{Res}[F_1(p)/F_2(p); p_k] = F_1(p)/F'_2(p)\Big|_{p=p_k}
\end{equation*}
The above analysis is the basis for running a residual machinery to calculate all the coefficients $\alpha _n(\tau )$.

\subsubsection{Residual machinery}

Let us denote via ${\cal I}_k$ the following contour integral
\begin{eqnarray*}\label{short}
{\cal  I}_k &=&   \dfrac{1}{2\pi \iu} \oint_{\gamma_k} \frac{\baru(p,\tau)} {\sinh\left(\sqrt{p} y(\tau) \right)} dp.
\end{eqnarray*}

Below we show that all coefficients $\alpha_n, \ n=1,\ldots, \infty$ can be expressed via these integrals.

\paragraph{ \bf{1. Coefficient $\alpha _1(\tau )$.}} Integrating \eqref{WInver2} along the contour $\gamma _1$ gives
\begin{align*} 
 \alpha_1(\tau ) e^{ - \left(\pi/y(\tau) \right)^2 \tau } \oint_{\gamma _1} \dfrac{1} {p + (\pi/y(\tau))^2}dp
 &+ \sum_{n=2}^{\infty} (-1)^{n+1} n \alpha_n(\tau) e^{ - \left( n \pi/y(\tau) \right)^2 \tau }
 \oint_{\gamma _1} \dfrac{1} { p + (n \pi/y(\tau))^2}dp  \\
 &= \dfrac{y(\tau)}{\pi} \oint_{\gamma _1} \dfrac{\baru(p,\tau)dp} {\sinh\left(\sqrt{p}y(\tau) \right)}. \nn
\end{align*}

Observe, that
\begin{equation*}\label{aRes}
\oint_{\gamma _1} \dfrac{1} { p + (\pi/y(\tau))^2}dp = 2 \pi \iu, \qquad \oint_{\gamma_1} \dfrac{1} { p + (n \pi/y(\tau))^2}dp = 0, \quad n \ge 2,
\end{equation*}
\noindent where the second result is due to the Cauchy integral theorem, \citep{cauchy1984}. Then
\begin{equation} \label{a1s}
 \alpha_1(\tau,y ) =  \frac{y(\tau)}{\pi}  e^{ \left( \pi/y(\tau) \right)^2 \tau } {\cal I}_1.
\end{equation}

\paragraph{ \bf{2. Coefficient $\alpha _2(\tau )$.}} By analogy, integrating the second equation of \eqref{WInver2} along the contour $\gamma _2$ we obtain
\begin{align*} 
\alpha_1(\tau)  & e^{-\left( \pi/y(\tau) \right)^2 \tau } \oint_{\gamma_2} \dfrac{dp}{p + (\pi/y(\tau))^2}
- 2\alpha_2(\tau) e^{-\left( 2\pi/y(\tau) \right)^2 \tau } \oint_{\gamma_2} \dfrac{dp}{p + (2 \pi/y(\tau))^2} \\
&+ \sum_{n=3}^\infty  (-1)^{n+1} \alpha_n(\tau) e^{ - \left(n \pi/y(\tau) \right)^2 \tau }
\oint_{\gamma_2}\dfrac{dp}{ p + (n \pi/y(\tau))^2} = \dfrac{y(\tau)}{\pi}
\oint_{\gamma_2} \dfrac{\baru(p,\tau)} {\sinh\left(\sqrt{p}y(\tau) \right)} dp, \nonumber
\end{align*}
\ni whence using again the residual theorem and \eqref{a1s} we find
\begin{equation*} \label{a2s}
 \alpha_2(\tau,y ) = -  \frac{y(\tau)}{2\pi} e^{ \left( 2\pi/y(\tau) \right)^2 \tau } \left[ {\cal I}_2 -  {\cal I}_1 \right].
\end{equation*}

\paragraph{ \bf{3. Coefficient $\alpha _n(\tau )$.}} Proceeding in a similar manner, we obtain a general formula for the coefficients
$\alpha_n, \ n \ge 1$
\begin{equation} \label{aNs}
 \alpha_n(\tau ) = (-1)^{n+1}   \frac{y(\tau)}{n\pi} e^{ \left( n \pi/y(\tau) \right)^2 \tau } \left[{\cal I}_n -  (1-\delta_{n,1}){\cal I}_{n-1} \right],
\end{equation}
\ni where $\delta_{n,1}$ is the Kronecker symbol.

\subsubsection{The final solution}

To calculate the integrals in the RHS of \eqref{aNs}, we rewrite them in the explicit form by using the solution for $\baru(p,\tau)$ previously found in \eqref{sol1}
\begin{equation*} \label{rhs}
{\cal I}_j =  \dfrac{1}{2\pi \iu}  \oint_{\gamma_k}  \frac{e^{p \tau}}{\sinh(y(\tau) \sqrt{p})} \left[ \int_0^\tau \Psi(s) e^{- p s} \sinh(y(s) \sqrt{p}) d s  +  \int_0^{y(0)} u(x,0) \sinh(x \sqrt{p}) dx \right] dp.
\end{equation*}
As $\sinh(x)$ is a periodic complex function with the period $\pi k /i$, the RHS of this equation is regular everywhere except simple poles , where ${\sinh\left(\sqrt{p}y(\tau) \right)}$ vanishes. It is easy to checks that these poles are exactly $ p_i, \ i=1,\ldots,k$. Therefore, we again can directly apply the Cauchy residual theorem.  Computing residuals, after some algebra we obtain
\begin{equation}\label{aAll}
\alpha_n(\tau) = \frac{2}{y(\tau)}  \Bigg[ \int_0^{y(0)} u(x,0) \sin \left(\frac{n \pi x}{y(\tau)}\right) dx +
\int_0^\tau e^{ \left( n \pi/y(\tau) \right)^2 s}\Psi(s)  \sin \left(\frac{n \pi y(s)}{y(\tau)}\right)  d s. \Bigg]
\end{equation}
Thus, from \eqref{WFour} and \eqref{aAll} we find the final solution
\begin{align*} 
u(x,\tau) &= \frac{2}{y(\tau)} \Bigg[ \sum_{n=1}^{\infty} e^{ - \left( \frac{n \pi}{y(\tau)}\right)^2 \tau } \sin\left(\dfrac{n\pi x}{y(\tau)} \right)
\int_0^{y(0)} u(x,0) \sin \left(\frac{n \pi x}{y(\tau)}\right) dx \\
&+ \sum_{n=1}^{\infty} \sin\left(\dfrac{n\pi x}{y(\tau)} \right)
\int_0^\tau e^{ \left( n \pi/y(\tau) \right)^2 (s-\tau)}\Psi(s)  \sin \left(\frac{n \pi y(s)}{y(\tau)}\right)  d s \Bigg]. \nonumber
\end{align*}
This can also be re-written as
\begin{align} \label{fin2}
u(x,\tau) &= \frac{2}{y(\tau)} \Bigg[ \int_0^{y(0)} dz \ u(z,0)
\sum_{n=1}^{\infty} e^{ - \left( \frac{n \pi}{y(\tau)}\right)^2 \tau }  \sin\left(\dfrac{n\pi x}{y(\tau)} \right) \sin \left(\frac{n \pi z}{y(\tau)}\right)  \\
&+ \int_0^\tau d s \ \Psi(s) \sum_{n=1}^{\infty}  e^{ \left( n \pi/y(\tau) \right)^2 (s-\tau)}  \sin \left(\frac{n \pi y(s)}{y(\tau)}\right) \sin\left(\dfrac{n\pi x}{y(\tau)} \right) \Bigg]. \nonumber
\end{align}
We proceed with the observation that the sums in \eqref{fin2} could be expressed via Jacobi theta functions of the third kind, \citep{mumford1983tata}. By definition
\begin{equation}\label{the3}
\theta_3 (z,\omega) = 1 + 2 \sum_{n=1}^{\infty} \omega^{n^2}\cos(2 n z).
\end{equation}
Therefore,
\begin{align} \label{sum1}
\sum_{n=1}^{\infty} & e^{ - \left( \frac{n \pi}{y(\tau)}\right)^2 \tau }  \sin\left(\dfrac{n\pi x}{y(\tau)} \right) \sin \left(\frac{n \pi z}{y(\tau)}\right) = \frac{1}{4}[\theta_3(\phi_-(z), \omega_1) - \theta_3(\phi_+(z),\omega_1)], \\
\sum_{n=1}^{\infty}  & e^{ \left( n \pi/y(\tau) \right)^2 (s-\tau)}  \sin \left(\frac{n \pi y(s)}{y(\tau)}\right) \sin\left(\dfrac{n\pi x}{y(\tau)} \right) = \frac{1}{4}[\theta_3(\phi_-(y(s)),\omega_2) - \theta_3(\phi_+(y(s)),\omega_2)], \nonumber \\
\omega_1 &= e^{ - \left( \frac{\pi \sqrt{\tau}}{y(\tau)}\right)^2}, \quad \omega_2= e^{ \left( \frac{\pi \sqrt{s-\tau}}{y(\tau)}\right)^2}, \quad  \phi_-(z) = \frac{\pi (x-z)}{2 y(\tau)},  \quad \phi_+(z) = \frac{\pi (x+z)}{2 y(\tau)}. \nonumber
\end{align}

 A well-behaved theta function must have parameter $|\omega| < 1$, \citep{mumford1983tata}. This condition holds at any $\tau > 0$.

Thus, \eqref{fin2} transforms to a simpler form
\begin{align} \label{fin2f}
u(x,\tau) &= \frac{1}{2 y(\tau)} \Bigg[ \int_0^{y(0)} dz \ u(z,0) [\theta_3(\phi_-(z), \omega_1) - \theta_3(\phi_+(z),\omega_1)]  \\
&+ \int_0^\tau d s \ \Psi(s) [\theta_3(\phi_-(y(s)),\omega_2) - \theta_3(\phi_+(y(s)),\omega_2)] \Bigg]. \nonumber
\end{align}
The RHS of \eqref{fin2} depends on $x$ via functions $\phi_-, \phi_+$. Since the theta function $\theta_3 (z,\omega)$ is even in $z$, the boundary condition at $x=0$ is satisfied. At $x=y(\tau)$ it is also satisfied as follows from \eqref{sum1} if one reads it from right to left.

The result in \eqref{fin2f} to some extent is not a surprise, as it is known that the Jacobi theta function is the fundamental solution of the one-dimensional heat equation with spatially periodic boundary conditions, \citep{Ohyama95}.

\section{Pricing American options}

We recall that an American option is an option that can be exercised at anytime during its life. American options allow option holders to exercise the option at any time prior to and including its maturity date, thus increasing the value of the option to the holder relative to European options, which can only be exercised at maturity. The majority of exchange-traded options are American. For a more detailed introduction, see \citep{Detemple2006, hull97}.

It is known that pricing American (or Bermudan) options requires solution of a linear complimentary problem. Various efficient numerical methods have been proposed for doing that. For instance, when the underlying stock price $S_t$ follows the time-dependent Black-Scholes model these (finite-difference) methods are discussed  in \citep{ItkinBook} (see also references therein).

Another approach, elaborated e.g., in \citep{Andersen2016} for the Black-Scholes model with constant coefficients, uses a notion of the exercise boundary $S_B(t)$. The boundary is defined in such a way, that, e.g., for the American Put option $P_A(S,t)$ at $S \le S_B(t)$ it is always optimal to exercise the option, therefore  $P_A(S,t) = K - S$. For the complimentary domain $S > S_B(t)$ the earlier exercise is not optimal, and in this domain $P_A(S,t)$ obeys the Black-Scholes equation. This domain is called the continuation (holding) region. The problem of pricing American options lies in the fact that $S_B(t)$ is not known in advance. Instead, we only know the price of the American option at the boundary. For instance, for the American Put we have $P_A(S_B(t),t)  = K - S_B(t)$, and for the American Call - $C_A(S_B(t),t)  = S_B(t) - K$. A typical shape of the exercise boundary for the Call option obtained with the parameters $K = 100, r = 0.05, q = 0.03, \sigma = 0.2$ is presented in Fig.~\ref{EB}. The method proposed in \citep{Andersen2016} finds $S_B(t)$ by numerically solving an integral (Volterra) equation for $S_B(t)$. The resulting scheme is straightforward to implement and converges at a speed several orders of magnitude faster than existing approaches.

In terms of this paper, the continuation region is a domain with the moving boundary where the option price solves the corresponding PDE. In case of our model in \eqref{OU1} this is the PDE in \eqref{PDE}. Therefore, this problem is, by nature, similar to that for the barrier options considered in Section~\ref{SecBO}, but the difference is as follows.
\begin{itemize}
\item For the barrier option pricing problem the moving boundary (the time-dependent barrier) is known, as this is stated in \eqref{bc}. But the Option Delta $\fp{u}{x}$ at the boundary $x = y(\tau)$ is not, and should be found by solving the linear Fredholm equation \eqref{fred}. Also, the problem is solved subject to the vanishing condition at the barrier (the moving boundary) for the option value.

\item For the American option pricing problem the moving boundary is not known. However, the option Delta $\fp{u}{x}$ at the boundary $x = y(\tau)$ is known (it follows from the conditions $\fp{C_A}{S}|_{S = S_B(t)} = 1$ and $\fp{P_A}{S}|_{S = S_B(t)} = -1$ expressed in variables $x$ and $y(\tau)$ according to their definitions in Section~\ref{SecBO}). Also the boundary condition for the American Call and Put at the exercise boundary (the moving boundary) differs from that for the Up-and-Out barrier option, namely: it is $C_A(S_B(t),t)) = S_B(t) - K$ for the Call, and $P_A(S_B(t),t)) = K - S_B(t)$ for the Put
\end{itemize}

Because of the similarity of these two problems, it turns out that the American option problem can be solved for the continuation region together with the simultaneous finding of the exercise boundary, by using the same approach that we proposed for solving the barrier option pricing problem. However, due to the highlighted differences, some equations slightly change.

\subsection{Solution of the American Call option pricing problem}

Since the PDE we need to solve is the same as in \eqref{PDE}, we do same transformations as in Section~\ref{SecBO}, and come up to the same heat equation as in \eqref{Heat}. It should be solved subject to the terminal condition
\begin{equation*} \label{tcA}
u(x,0) = \left(x e^{-\int_0^T w(s) ds} - K\right)^+ e^{-f(x,T)},
\end{equation*}
\noindent and the boundary conditions
\begin{align*} 
u(0,\tau) &= 0, \qquad u(y(\tau),\tau) \equiv \psi_1(\tau) = y(\tau) - K, \\
\Psi(\tau) &\equiv \fp{u}{x}\Bigg|_{x=y(\tau)} = \frac{e^{- f(y(\tau),t)}}{g^2(t)}\left[1 + a_1(t) y(\tau) (y(\tau) - K) \right],
\quad a_1(t) = \frac{g'(t)+g(t) (r(t)-q(t))}{g(t) \sigma (t)^2},  \nonumber
\end{align*}
\noindent where $t = t(\tau)$. We underline once again that the function $y(\tau)$ here is not known yet, while $\Psi(\tau)$ is known. These problems with the free boundaries are also well known in physics.

We proceed by using the same transformation  in \eqref{intTr}, and by analogy with \eqref{ode2} obtain the following Cauchy problem
\begin{align*} 
\fp{\baru}{\tau} &- p \baru = \Psi(\tau) \sinh(x \sqrt{p}) + \psi_1(\tau) \sqrt{p},  \\
\baru(p,0) &=  \int_0^{y(0)} u(x,0) \sinh(x \sqrt{p}) dx, \nonumber
\end{align*}
This problem can be solved explicitly to yield, \citep{kartashov2001}
\begin{equation*} \label{solA1}
\baru e^{- p \tau}  = \int_0^\tau \Psi(\tau) e^{- p \tau} \sinh(y(\tau) \sqrt{p}) d \tau  +  \int_0^{y(0)} u(x,0) \sinh(x \sqrt{p}) dx
+ \sqrt{p} \int_0^\tau e^{- p \tau} \psi_1(\tau) d \tau.
\end{equation*}
Accordingly, instead of \eqref{sol2} we obtain
\begin{align*} 
\int_0^\infty e^{- p \tau}  \left[\Psi(\tau) \frac{\sinh(y(\tau) \sqrt{p})}{\sqrt{p}} + y(\tau) \right] d \tau
&= \frac{K}{p} - \frac{1}{\sqrt{p}}\int_0^{y(0)} u(x,0) \sinh(x \sqrt{p}) dx   = \frac{K}{p} + \frac{F(p)}{\sqrt{p}}.
\end{align*}
This is a nonlinear Fredholm equation of the first kind, but now with respect to the function $y(\tau)$. It can also be solved numerically (iteratively).

The next step is to reduce our problem to that with homogeneous boundary conditions. This can be done by change of the dependent variable
\begin{equation*}
u(x,\tau) = W(x,\tau) + \Theta(x,\tau), \quad \Theta(x,\tau) = (1-x/y(\tau)] \psi_1(\tau).
\end{equation*}
The function $W(x,\tau)$ solves the same heat equation with the same terminal condition, and with the homogeneous boundary conditions. Therefore, it can be solved by using the method of generalized integral transform described in Section~\ref{GITmethod}. The solution reads
\begin{align*} 
u(x,\tau) &= \Theta(x,\tau) + \sum_{n=1}^{\infty} \alpha_n(\tau ) e^{ - \left( \frac{n \pi}{y(\tau)}\right)^2 \tau } \sin\left(\dfrac{n\pi x}{y(\tau)} \right), \\
\alpha_n(\tau)  &= \frac{2}{y(\tau)}\Bigg[ \int_0^{y(0)} [u(z,0) - \Theta(z,0)]\sin \left(\frac{n \pi z}{y(\tau)}\right) dz +
\int_0^\tau e^{ \left( n \pi/y(\tau) \right)^2 s} \left[\Psi(s) + \frac{\psi_1(s)}{y(s)}\right]  \sin \left(\frac{n \pi y(s)}{y(\tau)}\right)  d s \Bigg]. \nonumber
\end{align*}
Again, using the definition of the Jacobi theta function in \eqref{the3}, this can be finally re-written as
\begin{align*} 
u(x,\tau) &= \Theta(x,\tau) + \frac{1}{2y(\tau)}\Bigg[ \int_0^{y(0)} dz \ [u(z,0) - \Theta(z,0)] [\theta_3(\phi_-(z), \omega_1) - \theta_3(\phi_+(z),\omega_1)]  \\
&+  \int_0^\tau d s \ \left[\Psi(s) + \frac{\psi_1(s)}{y(s)}\right] [\theta_3(\phi_-(y(s)),\omega_2) - \theta_3(\phi_+(y(s)),\omega_2)] \Bigg]. \nonumber
\end{align*}

\section{Numerical example} \label{exper}

To test performance and accuracy of our method in this Section we provide a numerical example where a particular time dependence of $r(t), q(t), \sigma(t)$ is chosen as
\begin{equation} \label{ex}
r(t) = r_0 e^{-r_k t}, \qquad q(t) = q_0, \qquad \sigma(t) = \sigma_0 e^{-\sigma_k t}.
\end{equation}
Here $r_0, q_0, \sigma_0, r_k, \sigma_k$ are constants. With this model \eqref{Ric} can be solved analytically to yield
\begin{equation} \label{wSol}
w(t) = q_0 - r_0 e^{- r_k t}.
\end{equation}
Accordingly, from \eqref{gDef} we find
\begin{equation} \label{gSol}
g(t) = \exp \left[q_0 t + \frac{r_0}{r_k}  \left(e^{- r_k t} - 1 \right) \right],
\end{equation}
\noindent and from \eqref{sub1}
\begin{equation} \label{kT}
f(x,t) = k(t) = \frac{1}{2} \left[ \log\left( \frac{g(t)}{g(0)}\right)  - q_0 t + 3 \frac{r_0}{r_k} \left(1 - e^{- r_k t}\right) \right].
\end{equation}

The algorithm described in Section~\ref{SecBO} was implemented in python. We did it for two reasons. First, we found neither any standard implementation of the Jacobi theta functions in Matlab, nor any custom good one. Surprisingly, this is also not a part of numpy or scipy packages in python. However, they are available as a part of the python package mpmath which is a free (BSD licensed) Python library for real and complex floating-point arithmetic with arbitrary precision, see \citep{mpmath}. It has been developed by Fredrik Johansson since 2007, with help from many contributors.

Also, we didn't find any standard implementation of solver for the Fredholm integral equation of the first kind in both python and Matlab. Therefore, we implemented a Tikhonov regularization method as this is described in \citep{Fuhry2011}. In particular, with the model used in this Section, the function $F(p)$ reads
\begin{equation} \label{FpSol}
F(p) = \frac{e^{-k(T)}}{p} \left[-\sqrt{p} (K_1 - y_0) \cosh \left(\sqrt{p} y_0\right) + \sinh \left(K_1 \sqrt{p}\right) - \sinh \left(\sqrt{p} y_0 \right)\right].
\end{equation}

Finally to validate the results provided by our method, we implemented a FD solver for pricing Up-and-Out barrier options. This solver is based on the Crank--Nicolson scheme with a few Rannacher first steps, and uses a non-uniform grid, in more detail see, eg., \citep{ItkinBook}. We implemented two solvers: one for the backward PDE, and the other one - for the forward PDE. But logically, since in this paper we solved the backward PDE, it does make sense to compare our method with the backward solver. This implementation  has been done in Matlab.

In our particular test we choose parameters of the model as they are presented in Table~\ref{tab1}.

\begin{table}[!htb]
\begin{center}
\begin{tabular}{|c|c|c|c|c|c|c|c|c|}
\hline
$r_0$ & $q_0$ & $\sigma_0$ & $r_k$ & $\sigma_k$ & $H$ & $S_0$   \\
\hline
0.02 & 0.01 & 0.5$\cdot H$ & 0.1 & 0.2 & 90 & 60  \\
\hline
\end{tabular}
\caption{Parameters of the test.}
\label{tab1}
\end{center}
\end{table}
We recall, that here $\sigma(t)$ is the normal volatility. Therefore, we choose its typical value by multiplying the log-normal volatility by the barrier level.

\begin{figure}[!htb]
\vspace{-0.1in}
\begin{center}
\includegraphics[totalheight=3.5in]{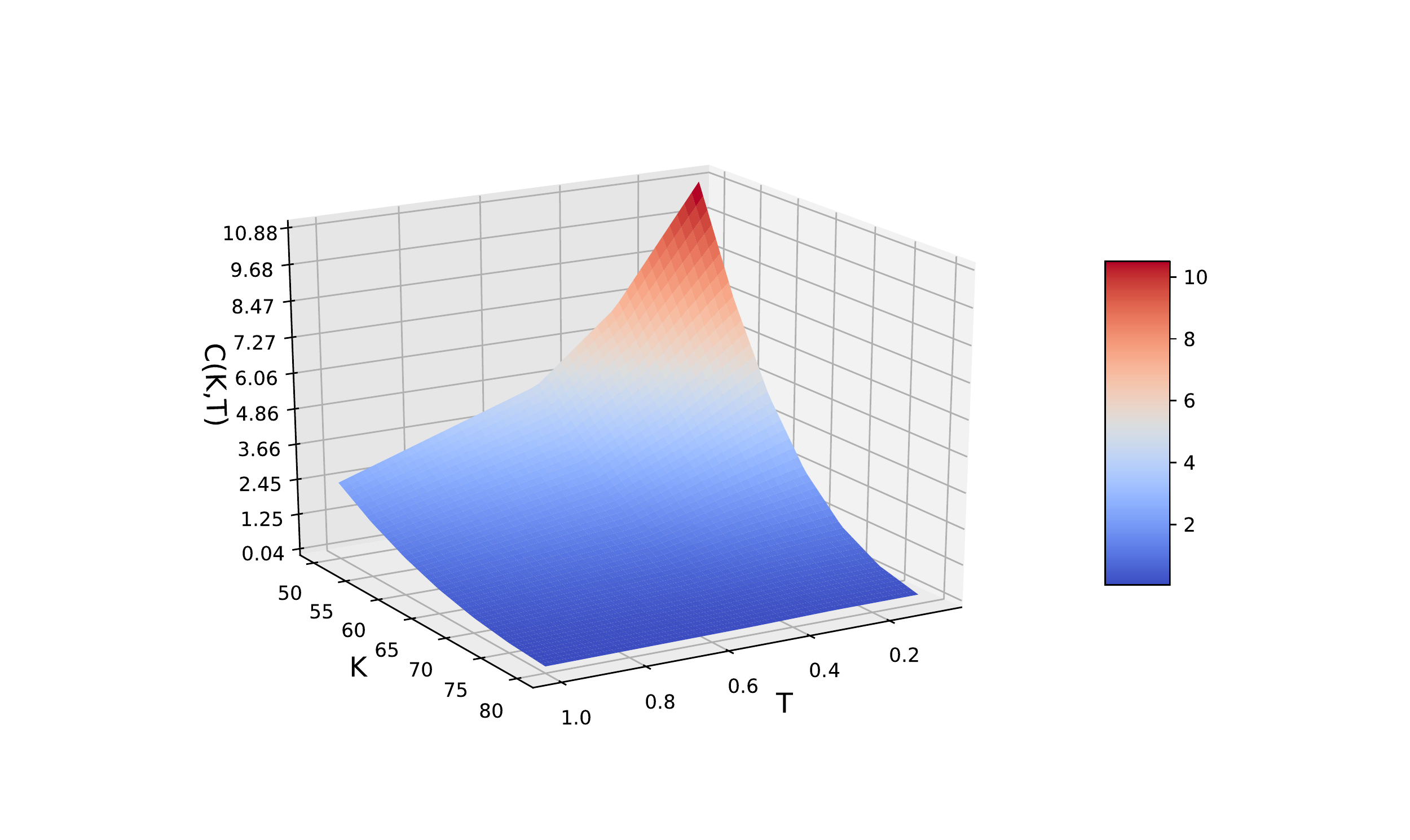}
\caption{Up-and-Out barrier Call option price computed by our method in the test.}
\label{figPrice}
\end{center}
\end{figure}

We run the test for a set of maturities $T \in [1/12, 0.3,0.5,1]$ and strikes $K \in [50, 55, 60, 65, 70, 75, 80]$. The Up-and-Out barrier Call option prices computed in such an experiment are presented in Fig.~\ref{figPrice}.

\begin{figure}[!htb]
\vspace{-0.1in}
\begin{center}
\includegraphics[totalheight=3.5in]{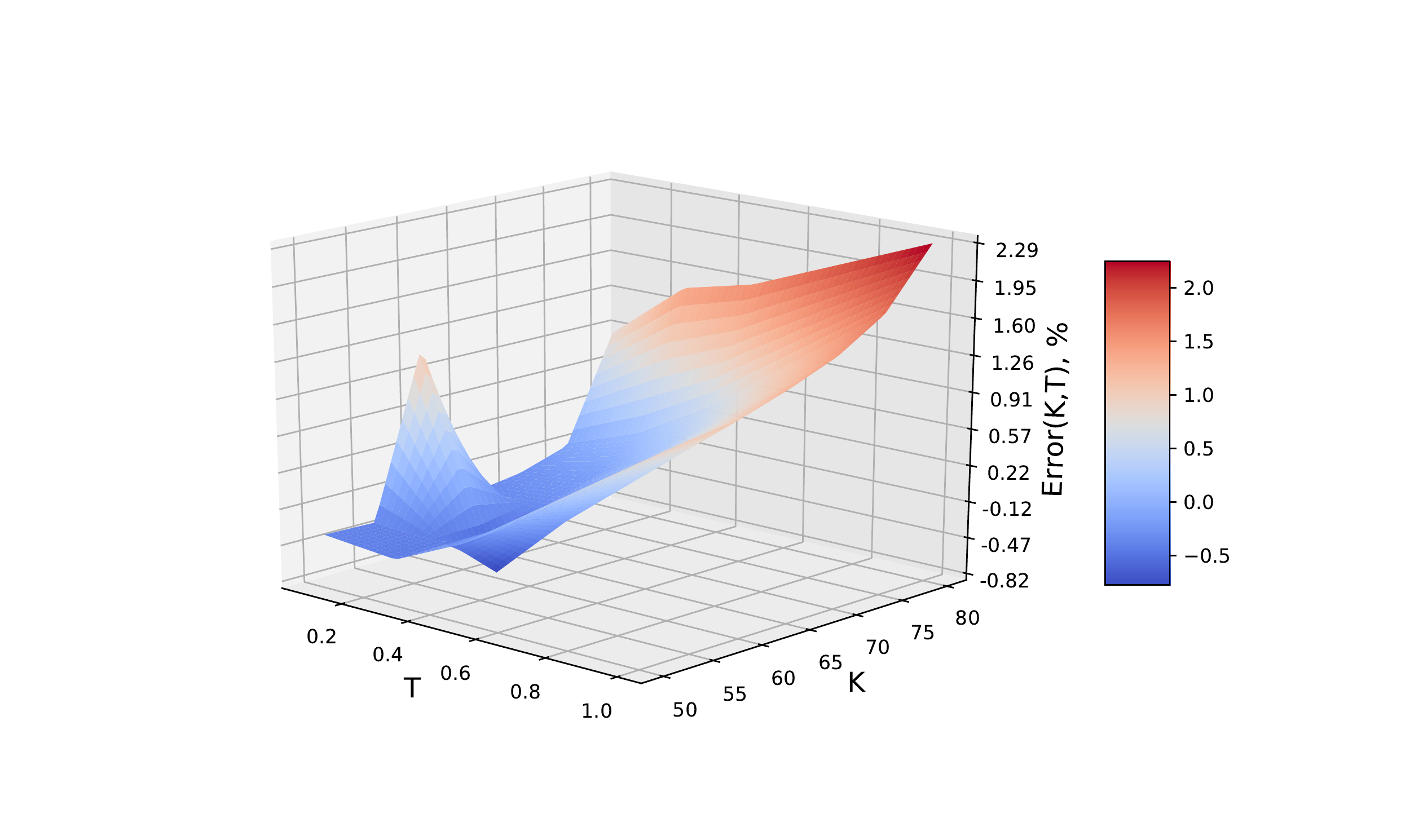}
\caption{The relative error in the Up-and-Out barrier Call option prices obtained by using  our method and the FD solver with 101 nodes in space $S$ and time step $\Delta t$ = 0.01.}
\label{error}
\end{center}
\end{figure}

In Fig.~\ref{error} the relative errors between the Up-and-Out barrier Call option prices obtained by using our method and the FD solver  are presented as a function of the option strike $K$ and maturity $T$. Here to provide a comparable accuracy we run the FD solver with 101 nodes in space $S$ and the time step $\Delta t$ = 0.01.

\begin{figure}[!htb]
\vspace{-0.1in}
\begin{center}
\includegraphics[totalheight=3.5in]{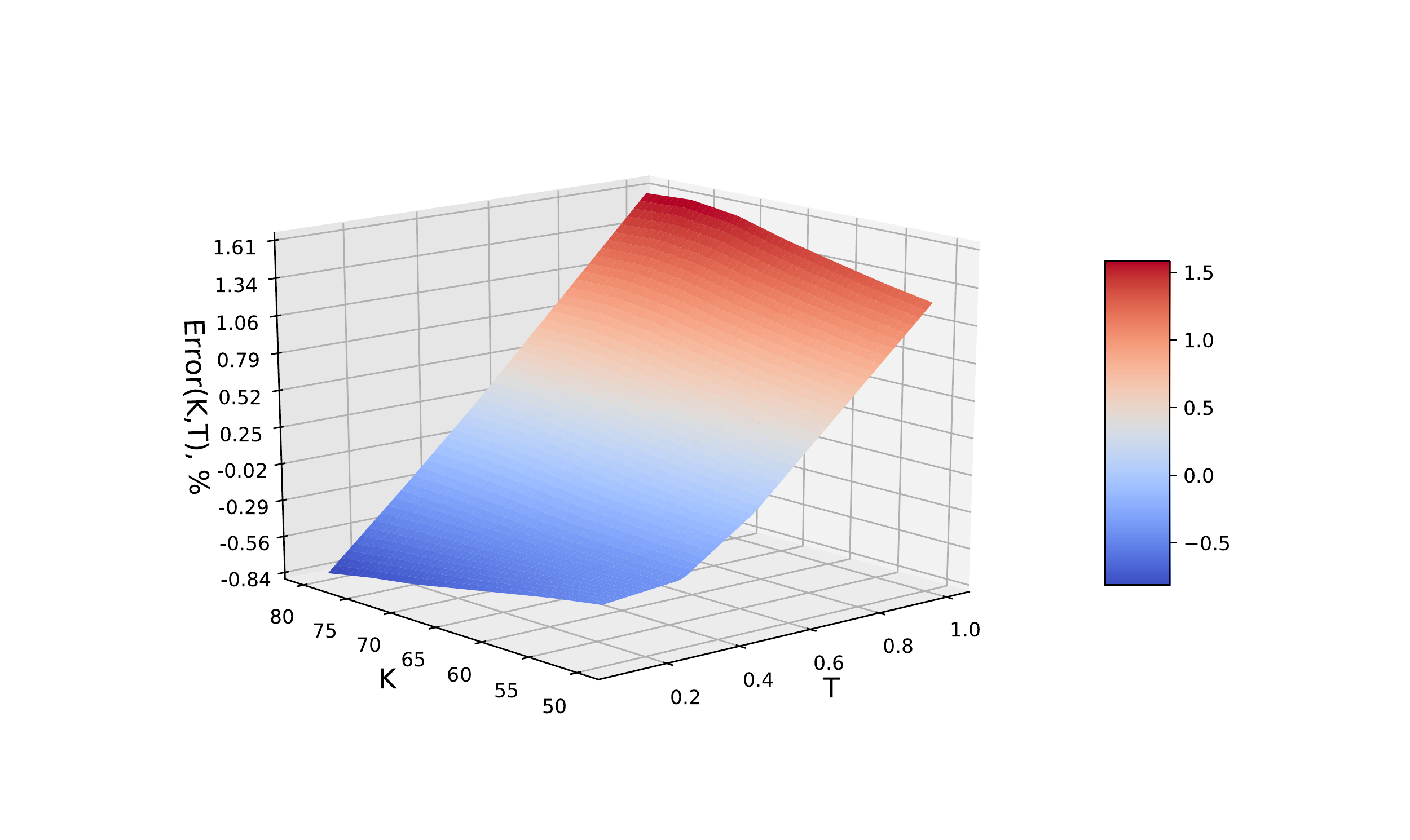}
\caption{The relative error in the Up-and-Out barrier Call option prices obtained by using  our method and the FD solver with 201 nodes in space $S$ and time step $\Delta t$ = 0.001.}
\label{error0001}
\end{center}
\end{figure}

It can be seen that the quality of the FD solution is not sufficient. Therefore, we reran it by using 201 nodes in space $S$ and the time step $\Delta t$ = 0.001. The relative errors between our semi-analytic and the FD solutions in this case are presented in Fig.~\ref{error0001}.

It can be seen that the agreement becomes better, so the relative error decreases. However, the cost for this improvement of the FD method is speed. In Table~\ref{elaps} we compare the elapsed time of both methods. The column "no $\Psi$" has the following meaning. Since the volatility and the interest rate change with time relatively slow, contribution of the second integral in \eqref{fin2f} to the option price is negligible. Therefore, in this particular case we can find the option price by computing only the first integral in \eqref{fin2f}. Accordingly, we don't need to solve the Fredholm equation \eqref{fred} that almost halves the elapsed time.

\begin{table}[!htb]
\small
\begin{center}
\begin{tabular}{| p{0.2\linewidth} | p{0.15\linewidth} | p{0.2\linewidth} | p{0.1\linewidth} | p{0.1\linewidth} | p{5em} | }
\hline
Test & semi-analytic & semi-analytic, no $\Psi$ & FD-101, $\Delta t=0.01$ & FD-201, $\Delta t=0.01$ & FD-201, $\Delta t=0.001$ \\
\hline
Elapsed time, sec & 0.23 & 0.16 & 0.36 & 0.65 & 3.6 \\
\hline
\end{tabular}
\caption{Elapsed time in secs of various tests.}
\label{elaps}
\end{center}
\end{table}
Finally, it is known that linear algebra in python (numpy) is almost  3 times slower than that in Matlab. Therefore, given the same accuracy, our method is about 30-40 times faster than the backward FD solver.

Of course, the forward FD solver is by an order of magnitude faster than the backward one if we need to simultaneously price multiple options of various strikes and maturities, but written on the same underlying. However, for barrier options this approach requires a very careful implementation, which often is not universal and with a lot of tricks involved.

\section{Discussion}

Our attention in Section~\ref{SecBO} was drawn to the Up-and-Out barrier Call option $C_{uao}$. Obviously, using the barriers parity, \citep{hull97}, the price of the Down-and-Out barrier Call option $C_{dao}$ can be found as $C_{dao} = C_{van} - C_{uao}$, where $C_{van}$ is the price of the European vanilla Call option. It is known that the latter is given by  the corresponding formula for the process with constant coefficients, where those efficient constant coefficients $\bar{r}, \bar{q}, \bar{\sigma}$ are defined as
\begin{equation*}
T \bar{r} = \int_0^T r(s) ds, \qquad T \bar{q} = \int_0^T q(s) ds, \qquad T \bar{\sigma}^2 = \int_0^T \sigma^2(s) ds.
\end{equation*}

Second, as shown in Section~\ref{GITmethod}, the barriers also could be some arbitrary functions of time, as this changes just the definition of function $y(\tau)$. And our method provides the full coverage of this case with no changes.

Third, and perhaps the most important point is about computational efficiency of our method.  In addition to what was presented in Section~\ref{exper}, let's look at this problem from a theoretical pint of view. Suppose the barrier pricing problem is
attacked by solving the forward PDE for a set of strikes $K_i, \ i=1,\ldots,\bar{k}$ and a set of maturities $T_j, \ j=1,\ldots,\bar{m}$ numerically by some FD method on a grid with $N$ nodes in the space domain $S \in [0,H]$, and $M$ nodes in the time domain $t \in [0,T_{\bar{m}}]$. Then the complexity of this method is known to be $O(MN + 4 N)$. This should be compared with the complexity  of our approach.

Let's assume that the Riccati equation in \eqref{Ric} can be solved either analytically, or, at least, approximately, as this is discussed  in Section~\ref{trHeat}. Then the first computational step consists in solving the linear Fredholm equation in \eqref{fred}. This can be done on a rarefied grid with $M_1 < M$ nodes and complexity $O(M_1^3)$. The intermediate values in $t$ can be found (if necessary) by interpolation with the complexity $O(M_1^2)$. As the integral kernel doesn't depend on strikes $K_i$, this calculation can be done simultaneously for all strikes still preserving the complexity $O(M_1^3)$.

The final solution of the pricing problem is provided in the form of two integrals in \eqref{fin2f}. Therefore, if we need the option price at a single value of $S_0$ (same as when solving the forward PDE), but for all strikes and maturities, the complexity is $O(2 \bar{k} L(M_1 + N_1)$, where $N_1$ is the number of points in the $x$ space, and $O(L)$ is the complexity of computing the Jacobi theta function $\theta_3 (z,\omega)$. Normally, $M_1 \le N, \ L \ll N, \ N_1 \ll N$ for the typical values of $N$ in the FD method (about 50--100 or even more). Thus, the total complexity of our method is fully determined by the solution of the Fredholm equation. Therefore, our method is slower than the corresponding FD method if $M_1 > (MN)^{1/3}$. For the American option this situation is worse since instead of solving a linear Fredholm equation we need to solve a nonlinear equation. This can be done iteratively, e.g., using $k$ iterations until the method converges to the given tolerance. Then the total complexity becomes $O(k M_1^3)$.  However, our experiments show that using just $M_1 = 10$ points in $p$ space could be sufficient, while further increase of $M_1$ doesn't change the results.

Also, the accuracy of the method in $x$ can be increased if one uses high-order quadratures for computing the final integrals. For instance, one can use the Simpson instead of the trapezoid rule that doesn't affect the complexity of our method. While increasing the accuracy for the FD method is not easy (i.e., it significantly increases the complexity of the method, e.g., see \citep{ItkinBook}).

\vspace{0.4in}


\newcommand{\noopsort}[1]{} \newcommand{\printfirst}[2]{#1}
  \newcommand{\singleletter}[1]{#1} \newcommand{\switchargs}[2]{#2#1}

\end{document}